\documentclass[12pt]{article}

\usepackage{natbib}
\usepackage{graphicx}
\usepackage{tikz}
\usepackage{amsmath,amsfonts,dsfont,url}
\usepackage{enumitem}
\usepackage[margin=1in]{geometry}

\newcommand{\bibsuffix}[1]{}
\newcommand{\mb}{\mathbf}

\newcommand{\mc}{\mathcal}

\newcommand{\ket}[1]{|#1\rangle}
\newcommand{\bra}[1]{\langle#1|}

\newcommand{\nrm}[1]{\| #1 \|}

\title{Relativistic Constraints on Interpretations of Quantum Mechanics}
\author{Wayne C. Myrvold \\ Department of Philosophy \\ The University of Western Ontario}
\date{In Eleanor Knox and Alastair Wilson, eds., \emph{The Routledge Companion to Philosophy of Physics}, Routledge, 2021.}

\begin{document}
\maketitle
\tableofcontents
\nocite{DieksVermaas}
\section{Introduction}  Though it is fairly noncontroversial that an empirically adequate quantum theory must be a quantum field theory, and must be able to treat of the relativistic regime of energies,   much of the literature on   interpretations of quantum theory has been   focussed on nonrelativistic quantum mechanics. There is some justification for this.  Nonrelativistic quantum mechanics is conceptually and mathematically simpler, and does, after all, work very well in a low-energy regime. Any solution of the measurement problem, or any account of the ontology of quantum theory, must yield sensible results in this regime, and so nonrelativistic quantum mechanics can serve as a testing-ground for the viability of interpretations.

It should not be assumed, however, that an approach to the interpretation of quantum mechanics will carry over straightforwardly  to the context of a relativistic quantum field theory.  There are  two sorts of challenges to be met.  One is adaptation of the approach to a theory with relativistic causal structure.  Another challenge arises from the fact that relativistic quantum theories are \emph{quantum field theories}, that is, quantum theories of systems with infinitely many degrees of freedom.

As we shall see below, each of these challenges can  pose a serious difficulty for an approach to the interpretation of quantum mechanics.  We will focus on four main avenues of approach: (i) additional beables theories, which include ``hidden-variables'' theories and modal interpretations, (ii) dynamical collapse theories, (iii) Everettian, or ``many worlds'' interpretations, and (iv) non-realist interpretations, which deny that quantum states represent anything in  physical reality independent of considerations of agents and their beliefs.

Summarizing briefly, the conclusions that will be drawn:

(i) Additional beables theories face serious difficulties with relativistic causal structure.  This can be dealt with by introducing a dynamically distinguished (albeit unobservable) foliation of spacetime, or else by making the additional beables relational, so that some quantities, such as pointer positions, that we think of as local beables, are not local beables after all.    Furthermore, modal interpretations that employ the spectral resolution of a system's quantum state to pick out the additional beables run into serious trouble when extended to quantum field theories, as they run the risk of picking out no non-trivial additional beables.

(ii) Dynamical collapse theories can be adapted, in a fairly straightforward way, to relativistic causal structure.  However,  a collapse  theory that respects the full spacetime symmetry of Minkowski spacetime and its causal structure can do so only at the price of introducing additional, nonstandard quantum degrees of freedom that are  unlike those that appear in familiar quantum field theories.

(iii) Everettian interpretations  fare better; they extend to the relativistic context with little or no adjustment.

(iv) Some non-realist interpretations bypass difficulties with relativity by limiting the scope of application of quantum mechanics.

\section{The Quantum Measurement Problem}
If a quantum theory is capable, in principle, of affording a complete description of the physical goings-on in the world (or even just the laboratory), then it must be possible to  model an experimental apparatus as a  complicated array of quantum systems, and the interactions of the apparatus with the system of interest as a process conforming to the laws of quantum dynamical evolution. This, however, poses a problem, as typical interactions will result in a quantum state that is a superposition of terms corresponding to distinct experimental outcomes.   The literature on what is (misleadingly) known as the interpretation of quantum mechanics deals with how to make sense of this.

Among approaches  that attempt a realist construal of quantum theories, we can distinguish three main avenues of approach.  One avenue accepts the usual, linear evolution of the quantum state, but does not saddle the quantum state with the burden of fully representing the physical world, but, rather, supplements it with additional structure.  These are traditionally called ``hidden-variables theories,'' but, as \citet{BellCosmo,BellQJ} has pointed out, this is misleading, since, according to a theory like this, it is via these additional variables that the world manifests itself.  This is the reason we are calling theories of this sort \emph{additional beables theories}.  These include the de Broglie-Bohm pilot wave theory, and also modal interpretations.

Another avenue of attack modifies the dynamics of the quantum state so that, in appropriate circumstances, a process closely approximating the textbook collapse of the state vector occurs.  This typically involves a stochastic modification of the dynamics, resulting in an indeterministic theory. The best-known theory of this type is the Ghirardi-Rimini-Weber (GRW) theory, on which collapses are instantaneous, discrete events; there is a continuous analog, known as \emph{continuous spontaneous localization} (CSL).

These two approaches modify the quantum formalism, either by adding additional structure, or by modifying the dynamics.  A third class of approaches retains the linear quantum dynamics, with its attendant superpositions of macroscopically distinct states,  and attempts to construe a quantum state description as a complete description.  These are \emph{Everettian}, or \emph{many-worlds} approaches to quantum theory.

An approach that does not fit neatly into these categories is the \emph{relational interpretation} advocated by  Carlo Rovelli. It is akin in some ways to Everett's original conception, which he called the \emph{relative-state} interpretation.  It differs from it in not taking quantum states to be representational. For more on this, see \citet{sep-qm-relational}, and references therein.

On any approach, empirical adequacy requires recovery of at least an approximation to the quantum probabilities for outcomes of experiments, insofar as these have been verified by experiment. This includes violations of  Bell inequalities. It might seem that this, by itself, leads to an unavoidable conflict with relativistic causality.  As we shall see, things are not so simple!

\section{Relativistic Spacetimes}
Compatibility with special relativity will be a concern in this chapter, and this amounts to the question of whether a theory can live happily in Minkowski spacetime, without adjoining any additional spacetime structure, such as a privileged relation of distant simultaneity.  But, of course, special relativity is false, and the spacetime in which we live and move and have our being is not Minkowski spacetime, but, as general relativity teaches us, a spacetime of variable and dynamic curvature.  Minkowski spacetime is of interest only as a local approximation, on scales at which curvature can be ignored.  And general relativity itself is, almost certainly, an approximation to a deeper theory that incorporates quantum gravity.

For this reason, we will be interested, not only in Minkowski spacetime, but in a wider class of relativistic spacetimes. The radical departure of special relativity from previous conceptions of spacetime lies in its rejection of  a privileged global temporal order.  In Galilean spacetime, for any event $p$, any other event is either in the past of $p$, the future of $p$, or is simultaneous with $p$, and the relation of simultaneity is, as one would expect, an equivalence relation.  In Minkowski spacetime, events that are spacelike separated are neither to the past nor to the future of each other, and, although there will always be some inertial reference frame whose accompanying Lorentz coordinates assign the two events the same $t$-coordinate, this is not simultaneity in any physically significant sense.

In Minkowski spacetime, define a relation $\prec$ that holds between spacetime points $p$, $q$, when there is a future-directed curve from $p$ to $q$ that is everywhere timelike or lightlike.  This relation is irreflexive (no point stands in that relation to itself), antisymmetric (if $p \prec q$ then it is not the case that $q \prec p$), and transitive.

Now define the relation $\sim$ of temporal unconnectedness, that holds between two points $p$ and $q$ if neither $p \prec q$ nor $q \prec p$. Distinct points that are temporally unconnected are said to be \emph{spacelike separated}.   This relation is reflexive and symmetric, but not transitive.  On the contrary, for any two points $p$, $q$ such that $p \sim q$, one can always find a third, $r$, such that $p \sim r$ but $r \prec q$.

All of this remains true in the curved spacetimes of general relativity, provided that they are sufficiently well-behaved.\footnote{There is no need here for a precise characterization of ``well-behaved,'' but global hyperbolicity suffices for our purposes.} This is a striking difference between relativistic spacetimes and Galilean spacetime.  In Galilean spacetime we have a relation $\prec$ of temporal precedence that is irreflexive and antisymmetric, but the corresponding relation of temporal unconnectedness is transitive, and hence spacetime can be uniquely be partitioned into equivalence classes under this simultaneity relation.

In this chapter, we will be concerned with spacetimes equipped with a relation $\prec$ of temporal precedence, from which we define a relation $\sim$ of temporal unconnectedness.  We will assume that $\prec$ is irreflexive and antisymmetric (no temporal loops).  We will say that  a spacetime's temporal structure is \emph{causally relativistic} iff  for any two points $p$, $q$ such that $p \sim q$, there exists a point $r$, such that $p \sim r$ and $r \prec q$. Note that this is a condition on causal structure and not a condition of symmetry under translations or boosts (which, of course, are additional conditions that one might impose).

On the usual notion of causation, a cause must temporally precede its effects, and so, if an event $p$ is a cause of an event $q$, we must have $p \prec q$.   It is this that motivates the claim that special relativity forbids action at a distance; without it, there is no reason to think that causal relations between spacelike separated events are incompatible with relativistic spacetime structure.   Given a causally relativistic spacetime, we will say that a theory is \emph{causally local} if all causal relations that the theory affords respect the condition that the cause be in the past of the effect and, hence, that there are no cause-effect relations between temporally unconnected events. There seems to be no harm in assuming that any spacetime point to the past of a point $q$ is potentially a locus of an event that is a cause of $q$.  If we make this assumption, then the relations of temporal precedence and potential causal influence coincide, and we may, as is usual in discussions of spacetime structure, treat temporal structure and causal structure interchangeably.

The condition of causal locality---that is, the condition that there be no cause-effect relations between spacelike separated points---is independent of symmetries such as boost symmetries. One can  invent theories that are causally local but not invariant under boosts, and one can invent theories  that violate the condition of causal locality without picking out a preferred reference frame.

There is a tradition of invoking retrocausality---that is, backwards-in-time causation---in explaining quantum phenomena.   This approach has its roots in a suggestion by O. Costa de Beauregard (\citeyear{Costa76}) and has been championed by a number of authors;  see \citet{sep-qm-retrocausality} for an overview.  The idea is to make causal connectability time-symmetric by permitting cause-effect relations in both temporal directions.  On this approach,  causal influences can propagate along time-like or light-like lines in either temporal direction.  Assuming that causal connectability is transitive, this turns  the relation of causal connectability into the  trivial relation that holds between any pair of spacetime points.  Any approach of this sort is, therefore,   straightforwardly causally nonlocal, in the sense we are using the phrase, as any two spacetime points, including those that outside of each other's lightcones,  are causally connected. As \citet[Ch. 6]{KastnerTransact} has argued, an approach of this sort is capable of respecting the symmetries of Minkowski spacetime.

\section{Quantum theories in relativistic spacetimes}
A quantum-mechanical theory is what results when one subjects a classical theory of a system  with finitely many degrees of freedom (such as, for example, a finite number of particles) to quantization. A quantum field theory, on the other hand, results from quantization of a classical theory with infinitely many degrees of freedom. Introductory quantum field theory textbooks typically contain, in their early chapters, arguments that a relativistic quantum theory must be a quantum field theory.  For a precise version of one such argument, see \citet{DefenseofDogma}.

There are, therefore, two sorts of difficulties one might encounter when one seeks to extend  to the relativistic domain an approach to the quantum measurement problem that has been formulated in terms of nonrelativistic quantum mechanics.  One  source of difficulties  might lie in a adapting the theory to relativistic causal structure.  Another source of difficulties might stem from differences between quantum mechanics and  quantum field theories.

Typically, relativistic quantum field theories are expressed in the Heisenberg picture, rather than the Schr\"odinger picture.  The Schr\"odinger picture, or an analogue of it, can, however, be formulated in a  relativistic spacetime.   On this picture, we track the rate of change of an observable via a changing state vector.  Take a foliation of spacetime into spacelike hypersurfaces (most conveniently, hyperplanes, but we can also consider nonflat surfaces), taken to be smooth, so that there is, at any point of each of these hypersurfaces, a vector normal to the surface. Integral curves of this vector field form a timelike congruence. Pick one hypersurface $\sigma_0$,  and consider a field operator $\hat{\phi}(x)$. For each $x_0$ on $\sigma_0$, there will be a family of operators $\hat{\phi}(x)$ corresponding to points on the integral curve containing $x_0$, and, for any quantum state, one can consider how the expectation value of $\hat{\phi}(x)$ changes as one follows these timelike curves.  This can be expressed, either in the Heisenberg picture, using the same state vector and different operators $\hat{\phi}(x)$ for different  points on an integral curve, or in the Schr\"odinger picture, using the same operator $\hat{\phi}(x_0)$ for every spacetime point on the integral curve containing $x_0$, and different state vectors for evaluating expectation values of observables on different elements of our foliation.

There is also an analog of the interaction picture, called the \emph{Tomonaga-Schwinger picture}.  On this picture, one writes the Lagrangian density as a sum of a free-field Lagrangian and a term incorporating interactions:
\begin{equation}
\mc{L}(x) = \mc{L}_0(x) + \mc{L}_I(x).
\end{equation}
The operators employed are solutions of the free-field equations, and one associates, with each spacelike Cauchy surface $\sigma$ (whether flat or not), a state vector $\ket{\psi(\sigma)}$.  Evolution from a surface $\sigma$ to another, $\sigma'$,  differing by an infinitesimal  deformation about a point $x$, satisfies the \emph{Tomonaga-Schwinger equation} \citep{Tomonaga46,Schwinger48}:
\begin{equation}
i \hbar  c\, \frac{\delta \ket{\psi(\sigma)}}{\delta \sigma(x)} = \mc{H}_I(x) \ket{\psi(\sigma)}.
\end{equation}
Integration of this equation yields, for any Cauchy surfaces $\sigma$, $\sigma'$, a  unitary mapping from $\ket{\psi(\sigma)}$ to $\ket{\psi(\sigma')}$.  This approach is not commonly discussed in recent textbooks but can be found in some older texts; see, \emph{e.g.}, \citet{SchweberQFT}.

\section{Additional-Beables Theories}

\subsection{De Broglie-Bohm theory}
The best-known additional-beables theory is the de Broglie-Bohm pilot wave theory.  On this theory, ordinary physical objects consist of point particles, or corpuscles, whose motion is guided by the quantum wave function.   The dynamical  laws of theory are the Schr\"odinger equation for the evolution of the quantum wave function, and the guidance equation, which fixes the velocities of Bohmian corpuscles.
\begin{equation}\label{guidance}
m_i \mb{v}_i(\mb{x}_1, \mb{x}_2, \ldots, \mb{x}_n, t) = \boldsymbol{\nabla}_i S(\mb{x}_1, \mb{x}_2, \ldots, \mb{x}_n, t),
\end{equation}
where $S/\hbar$ is the phase of the wave function:
\begin{equation}
S(\mb{x}_1, \mb{x}_2, \ldots, \mb{x}_n, t) = \hbar \: \mbox{Im} \: \log \Psi(\mb{x}_1, \mb{x}_2, \ldots, \mb{x}_n, t).
\end{equation}
It is assumed that, at some time $t_0$, the probability density for configurations of the corpuscles is given by the square of the wave function amplitude.  If this holds for some time $t_0$, the dynamics, consisting of the Schr\"odinger equation and the guidance equation, ensures that it holds for all time.  This property is known as  \emph{equivariance} of the Born-rule distribution.

\subsubsection{Causal structure}\label{BohmCausal}
It can be seen from the guidance equation (\ref{guidance}) that the velocity of any one of the corpuscles depends on an instantaneous configuration, and may, in principle, depend on the positions of arbitrarily many corpuscles, arbitrarily far away. Thus, the guidance equation  requires a distinguished relation of distant simultaneity for its formulation.

As an illustration, consider a pair of spin-$1/2$ particles in the singlet state, spatially separated and sequentially subjected to Stern-Gerlach experiments for the same direction, say the $z$-direction, which we will regard as the vertical direction.   The quantum state can be written,
\begin{equation}
\ket{\Psi(t)} = \psi_{+-}(\mb{x}_1, \mb{x}_2,t) \: \ket{z^+}_1\ket{z^-}_2 + \psi_{-+}(\mb{x}_1, \mb{x}_2,t) \: \ket{z^-}_1\ket{z^+}_2
\end{equation}
Before the experiment, there is no correlation between position and spin-state, and the two wavepackets overlap completely.
\begin{equation}
\psi_{+-}(\mb{x}_1, \mb{x}_2,t_0) = \psi_{-+}(\mb{x}_1, \mb{x}_2,t_0).
\end{equation}
The outcome of both experiments is determined by the position of the particle on which the experiment is done first.   The Stern-Gerlach experiment on particle 1 separates the wavepackets $\psi_{+-}$ and $\psi_{-+}$ along the $z_1$-direction.  Since Bohmian trajectories cannot cross, there will be a plane or other two-dimensional surface in $\mb{x}_1$-space such that particle 1 will  go in the $+$ direction if it is initially located above this plane, in the $-$ direction, if it is located below.  After this experiment, there is no longer any overlap, in configuration space, between the two wavepackets, and the two particles are  determinately located in only one of them.  The second  particle responds to its Stern-Gerlach apparatus accordingly, responding $+$ or $-$ depending on which wavepacket it is in.

This means that, if the experiment  on  particle 1 is done first, the outcome of both experiments is determined by the initial position of particle 1. By parity of reasoning, if the experiment on particle 2 is done first, the outcome of both experiments is determined by the initial position of particle 2.  These outcomes could differ, meaning that, for the same initial wavefunction and initial configuration of the particles, the outcomes of the experiments could differ depending on which experiment is done first.   Thus, the theory requires there to be a matter of fact about which experiment is done first, even if the experiments are done at spacelike separation.

We might ask: is this an artefact of the way this theory is formulated, or is it inherent in the very project of formulating a theory of this type, on which particles have definite trajectories and probabilities over configurations are to be given by the Born rule, with dynamics for the particles that ensures that this probability distribution is equivariant?

As \citet{Berndl96} have shown, the answer to this question is that a theory of this type must, indeed, invoke a privileged foliation, as it is impossible for the distribution postulate to be satisfied on all spacelike hyperplanes.

 Suppose that we have a pair of spin-$1/2$ particles, initially both located in the same  small region.  The particles are then separated a large distance, and each is passed through a Stern-Gerlach apparatus oriented in the $x$-direction, which separates the wavepacket (in configuration space) along the   $x$-axis of the particle it's interacting with. The beams are then carefully recombined, with care taken not to let the particles interact with anything that might induce decoherence, and then are passed through Stern-Gerlach devices oriented in the $z$-direction.  We suppose that all of this splitting,  recombining, and resplitting of the two beams occurs at spacelike separation.

We consider four hypersurfaces (see Figure \ref{planes}):
\begin{enumerate}
\item A hypersurface $\alpha$, on which both particle 1 and particle 2 have their wavepackets split into $x+$ and $x-$ wavepackets.
\item A hypersurface $\beta$, on which both particle 1 and particle 2 have their wavepackets split into $z+$ and $z-$ wavepackets.
\item A hypersurface $\gamma$, on which the wavepackets are split along $z_1$ and $x_2$ axes, respectively.
\item A hypersurface $\delta$, on which the wavepackets are split along $x_1$ and $z_2$ axes, respectively.
\end{enumerate}
\setlength{\unitlength}{0.025cm}
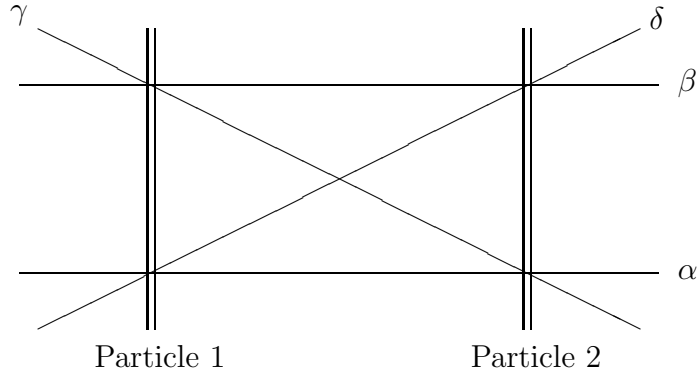
\begin{figure}[h]
\begin{centering}
\begin{picture}(360, 270)
\put(0,80){\line(1,0){340}}
\put(0,180){\line(1,0){340}}
 \put(10,50){\line(2,1){320}}
\put(10,210){\line(2,-1){320}}
\put(68,50){\line(0,1){160}}\put(72,50){\line(0,1){160}}
\put(268,50){\line(0,1){160}} \put(272,50){\line(0,1){160}}
\put(40,30){Particle $1$}
\put(240,30){Particle $2$}
 \put(350,77){$\alpha$} \put(350,177){$\beta$}
\put(-5,215){$\gamma$}\put(335,210){$\delta$}
\end{picture}\caption{The hypersurfaces considered in the proof.}\label{planes}
\end{centering}
\end{figure}
It is not, in fact, always possible to satisfy the distribution postulate along all four of these hypersurfaces.    Suppose that our pair of spin-$1/2$ particles is  prepared in the Hardy state:
\begin{eqnarray}
\ket{\psi}_{H} &=&
\frac{1}{\sqrt{12}}\left(\ket{x^+}_1 \ket{x^+}_2 - \ket{x^+}_1 \ket{x^-}_2 - \ket{x^-}_1 \ket{x^+}_2 - 3 \ket{x^-}_1 \ket{x^-}_2  \right) \label{alpha}
\\
&=&  \frac{1}{\sqrt{3}}\left(\ket{z^+}_1 \ket{z^-}_2  - \ket{z^+}_1 \ket{z^+}_2 + \ket{z^-}_1 \ket{z^+}_2 \right) \label{beta}
\\
&=&  \frac{1}{\sqrt{3}}\left( \ket{z^-}_1 \ket{z^+}_2    - \sqrt{2} \: \ket{z^+}_1 \ket{x^-}_2\right) \label{gamma}
\\
&=& \frac{1}{\sqrt{3}}\left( \ket{z^+}_1 \ket{z^-}_2    - \sqrt{2} \: \ket{x^-}_1 \ket{z^+}_2\right) \label{delta}
\end{eqnarray}
Let us write the states on these four hypersurfaces, with the position degrees of freedom included, writing $\ket{+}$ for the particle being in the $+$ beam, $\ket{-}$ for its being in the $-$ beam.
\begin{multline}
\ket{\psi(\alpha)} = \frac{1}{\sqrt{12}}\left(\ket{x^+}_1 \ket{+}_1 \ket{x^+}_2 \ket{+}_2  - \ket{x^+}_1 \ket{+}_1  \ket{x^-}_2 \ket{-}_2 \right.
\\
\left. - \ket{x^-}_1 \ket{-}_1 \ket{x^+}_2 \ket{+}_1 - 3 \ket{x^-}_1 \ket{-}_1  \ket{x^-}_2 \ket{-}_2   \right)
\end{multline}
\begin{equation}
\ket{\psi(\beta)}  = \frac{1}{\sqrt{3}}\left(\ket{z^+}_1 \ket{+}_1 \ket{z^-}_2 \ket{-}_1  - \ket{z^+}_1 \ket{+}_1  \ket{z^+}_2 \ket{+}_2
 + \ket{z^-}_1 \ket{-}_1  \ket{z^+}_2 \ket{+}_1  \right)
\end{equation}
\begin{equation}
\ket{\psi(\gamma)}  = \frac{1}{\sqrt{3}}\left( \ket{z^-}_1  \ket{-}_1  \ket{z^+}_2 \ket{+}_2    - \sqrt{2} \: \ket{z^+}_1 \ket{+}_1  \ket{x^-}_2 \ket{-}_2  \right)
\end{equation}
\begin{equation}
\ket{\psi(\delta)} = \frac{1}{\sqrt{3}}\left( \ket{z^+}_1  \ket{+}_1 \ket{z^-}_2  \ket{-}_2    - \sqrt{2} \: \ket{x^-}_1  \ket{-}_1  \ket{z^+}_2  \ket{+}_2 \right)
\end{equation}
 If the distribution postulate is satisfied on $\alpha$, there is a probability $1/12$ that both particles are in their respective $x^+$ beams. If the postulate is satisfied on $\gamma$, then in such cases particle 1 must be in its $z^-$ beam on $\gamma$. If  the postulate is satisfied on $\delta$, then  particle 2 must be in its $z^-$ beam on $\delta$.   But this means that, on $\beta$,   both particles will be in the $z^-$ beams, which should have probability zero. Conclusion:  the distribution postulate cannot be satisfied along all foliations, and the dynamics for the particles must be choosy about which, if any, foliation, it is with respect to which equivariance is to be maintained.

The argument does not depend on preparing the Hardy state exactly, or that our theory yield quantum probabilities exactly.  On a theory of this sort, there must be a joint probability distribution over the positions of the particles at each point of intersection of the hyperplans considered  that yields the correct correlations on each of the four hyperplanes.  As demonstrated by \citet{FineJointa,FineJointb},  violation of a Bell-type inequality for these variables is equivalent to impossibility of a joint distribution of that sort.

For example: if there is a joint distribution over these variables, we will have to have,
\begin{equation}
Pr(x_1^+(\alpha), x_2^+(\alpha)) \leq  Pr(x_1^+(\delta), z_2^+(\delta)) + Pr(z_1^+(\gamma), x_2^+(\gamma)) + Pr(z_1^-(\beta), z_2^-(\beta)),
\end{equation}
which is violated by the Hardy state.

A trajectory theory must choose a foliation along which the distribution postulate is to be satisfied. Moreover, it is fairly easy to see that, in order to maintain the distribution along this foliation, the dynamics must permit causal dependence of the behaviour of particles on choice of experiment at spacelike separation.

Consider three scenarios.
\begin{enumerate}[label=\Alph*.]
\item For particle 1, operations proceed as above, recombining the beams and resplitting them into $z^+$ and $z^-$ wavepackets. Nothing is done to particle 2 between $\alpha$ and $\beta$, and there is a  split into $x^+$ and $x^-$ wavepackets for particle 2 on both $\alpha$ and $\beta$.
\item Nothing is done to particle 1 between $\alpha$ and $\beta$, and it remains split into $x^+$ and $x^-$ wavepackets on both $\alpha$ and $\beta$.  Recombining and resplitting proceed as above, for Particle 2.
\item Recombining and resplitting proceed as above, for both particles.
\end{enumerate}
Suppose that we have dynamics according to which, if the probabilities for positions of particles match the Born-rule probabilities on $\alpha$, they do so on $\beta$, for all three scenarios.  Suppose, also, in scenarios $A$ and $B$, the particle to which nothing is done does not hop from one of its wavepackets to another. That is, in scenario $A$, if particle 2 is in the $x^+$ beam on $\alpha$, it is in the $x^+$ beam on $\beta$.

These conditions can be satisfied only if, for some initial configurations on $\alpha$, transition probabilities---that is probabilities of a particle ending up in a certain position on $\beta$, given the configuration on $\alpha$---for one particle depend on what is being done to the other particle.

To see this, suppose the contrary.  Suppose that, for any configuration on $\alpha$, the probability that particle 1 ends up in the $z^+$ beam on $\beta$ is the same, whether scenario $A$ or $C$ is in effect, and that the probability that particle 2 ends up in the $z^+$ beam on $\beta$ is the same, whether scenario $B$ or $C$ is in effect.  Then, since, on Scenario $A$, particle 2 is in the same beam on $\alpha$ and $\beta$, if the Born rule is satisfied on $\beta$, it is also satisfied on $\gamma$.  If transition probabilities for particle 1 are the same in scenarios $A$ and $C$, it follows that, in scenario $C$, the Born rule is satisfied  on $\gamma$.

Running the same argument with the roles of particles 1 and 2 reversed yields the conclusion that, if in scenario $B$ the Born rule is satisfied on $\beta$, and if transition probabilities for particle 2 are the same in scenarios $B$ and $C$, it follows that, in scenario $C$, the Born rule is satisfied  on $\delta$.  But, as we have seen, in scenario $C$, the Born rule cannot be satisfied on all four hypersurfaces.

\subsubsection{Bohmian quantum field theories} Even though there are good reasons to think that an additional-variables theory must employ a preferred foliation, the question still arises whether there could be a theory of this sort that recovers the empirical content of relativistic quantum field theories.  This would, of course, entail that the preferred foliation is empirically inaccessible.  Two avenues of approach suggest themselves for extending Bohmian theories to the context of quantum field theories. One is to retain a basic particle ontology, and permit particle creation and annihilation.  The other is to take classical field configurations as the basic ontology, and to provide a non-classical dynamics for the evolution of these field configurations.  For an overview of these options, see \citet{StruyveBohmianQFT}.  Struyve concludes that the field approach works best for bosonic fields, but is problematic when one attempts to extend it to fermionic fields.  For fermionic fields, the particle approach seems more appropriate.

The task of extending additional-beables theories to the context of quantum field theories remains a work in progress.  However, there do not seem to be in-principle obstacles to the creation of a theory of this sort that recovers the phenomena that are currently handled by quantum field theories.

\subsection{Modal interpretations}  Modal interpretations  supplement the quantum state with additional beables.  A variety of interpretations along these lines have been proposed; see \citet{sep-qm-modal} for an overview. Many of these share the feature that the nature of the additional beables depends on the quantum state. Given a quantum system $\alpha$, which may be a subsystem of a larger system, one considers the reduced state of $\alpha$, obtained by tracing out the degrees of freedom of the rest of the system.  If the state of the larger system is a pure quantum state, the reduced state of $\alpha$ will be pure if $\alpha$ is not entangled with the rest of the system, mixed, if it is. One resolves the density operator representing the reduced state into its spectral components:
\begin{equation}
\rho_\alpha = \sum_i w_i\: P_i,
\end{equation}
where $\{P_i\}$ are mutually orthogonally projections.  In the non-degenerate case, in which all nonzero coefficients $\{w_i\}$ are distinct, this resolution is unique.  The \emph{basic modal rule}  assigns definite values 1 or 0 to these projections; one of these projections represents a property definitely possessed by the system, the others, definitely absent.  The probability that $P_i$ is the possessed property is $w_i$.

\subsubsection{Difficulties arising from relativistic causal structure} Can  modal interpretations respect relativistic causal structure, or do they require a distinguished relation of distant simultaneity?   Dickson and   Clifton (\citeyear{DicksonCliftonModalRel})  proved, for  a  broad class of modal interpretations, that is impossible to furnish the theory with transition probabilities in such a way as to yield Born-rule probabilities for possessed values on arbitrary hyperplanes.  Dickson and Clifton's proof relies on an assumption concerning the transition probabilities for possessed values, the assumption they call ``stability,'' but, as \citet{ArntzCurious}  pointed out, the stability requirement is dispensable and the core of the proof concerns the nonexistence of certain joint distributions
yielding the appropriate Born probabilities as marginals.  Taking these proofs as a starting point, \citet{MR,CC}  adapted the argument of \citet{Berndl96}, outlined above, to show that any modal interpretation, so long as the possessed values of a system are capable of being regarded as local beables, cannot have Born-rule probabilities for these possessed values satisfied on arbitrary spacelike hyperplanes. The argument at the end  section \ref{BohmCausal}, above, can be straightforwardly adapted to show that these theories, also,  must  permit causal dependence of the behaviour of the additional beables on choice of experiment at spacelike separation.

Undeterred, Berkovitz and Hemmo(\citeyear{BerkHemmoPS2005},\citeyear{BerkHemmoFooP2005},\citeyear{BerkHemmo2006}) have proposed a relational modal interpretation.   The interpretation assigns properties to systems that are not thought to be intrinsic to that system, but are possessed by it only relative to some other system. In a relativistic context, these relational values may depend on what hypersurface is used to define a global state of the universe, and it can happen, in the presence of entanglement at spacelike separation, that some property of a system, at a single position on its worldline,  may have different values relative to different hypersurfaces.  Thus, the possessed values are not to be thought of as local beables, and, as the authors argue in some detail, the no-go theorems don't apply.

\subsubsection{Difficulties arising from infinitely many degrees of freedom}  In addition to the difficulties for modal interpretations arising from relativistic causal structure, there is another difficulty for attempts to extend modal interpretation to the context of relativistic quantum field theories, arising from infinitude of degrees of freedom.

There is, first, the issue of extending the basic modal rule to the context of a relativistic quantum field theory.  We want to be able to apply the rule to local subsystems of the universe.  If we consider a bounded region $R$ of spacetime, and the algebra of operators $\mc{A}(R)$ formed from fields in $R$, there is a difficulty in applying the basic modal rule, since, as pointed out by \citet{DieksModalRel}, these   local algebras will not contain either a density operator representing the state, or projections onto finite-dimensional subspaces.  \citet{CliftonModalRel} provided a version of the modal rule that makes sense in the context of algebraic quantum field theory, and also showed that there exists a whole host of states for which the rule picks out no definite observables other than multiples of the identity (that is, observables that have the same value no matter what the state is).   \citet{EarmanRuetscheModalRel} extend  Clifton's result, and discuss the  significance of these results for modal interpretations of quantum theory.

These trivialization results mean that a modal interpretation of this sort cannot respect relativistic causal structure and symmetries  while accomplishing the goal of providing a solution to the measurement problem. The purpose of a modal interpretation is  to have observables such as apparatus pointer observables take on definite values in physically realistic states.  Obviously, this is not accomplished if no observables are definite other than multiples of the identity.

\section{Collapse Theories} A dynamical collapse theory modifies the evolution of the quantum state. One approach that might be tried would be to replace the deterministic, unitary evolution of the quantum state by a deterministic and non-unitary evolution.  There are arguments to the effect that, given plausible assumptions,  this would  lead to the possibility of superluminal signalling \citep{Gisin89,SBGNoSignalling}.\footnote{These assumptions may be questioned; see  \citealt{KentNonlinear}.}  For this reason, in  this section we focus on stochastic collapse theories.

On a theory such as this, the dynamical laws do not specify a unique future state, given the state at a time $t_0$; rather, which state will be realized is, on such a theory, a matter of chance.  The idea is to specify a probabilistic law that approximates the usual unitary, deterministic evolution within those domains where we have good evidence for it, but departs sufficiently from it in some situations---which should include experiments that lead to recorded outcomes---so as to closely approximate the textbook collapse postulate, according to which some definite outcome obtains at the end of an experiment, with probabilities as to which outcome is obtained given by the Born rule.  One theory of this sort, with unitary evolution punctuated by discrete jumps, is the Ghirardi-Rimini-Weber (GRW) theory, referred to by the authors as \emph{Quantum Mechanics with Spontaneous Localization} (QMSL) \citep{GRW}.  Another is the \emph{Continuous Spontaneous Localization} (CSL) theory (\citealt{Pearle1989}, Ghirardi, Pearle, and Rimini \citeyear{GPR90}).

Among the experiments to be accounted for are tests of Bell inequalities. Since our theory is required to produce correct probabilities for the outcomes of such experiments, we do not expect the probabilistic law to be one on which spacelike separated events are probabilistically independent.  It might seem obvious, at first, that this dooms compatibility with relativity from the start.

Things are not so simple, however. Bell  himself said that, when he first encountered the GRW theory, ``I thought I could blow it out of the water, by showing it was \emph{grossly} in violation of Lorentz invariance'' \citep[p. 17]{BellExactQM}. A careful examination led him to the opposite conclusion, namely, that, though couched in terms of nonrelativistic quantum mechanics, the theory exhibits ``a residue, or at least an analogue, of Lorentz invariance'' (\citealt[p. 10]{BellQJ} in \citealt[pp. 206--207]{BellBook}).  The idea is this.  On the usual, Schr\"odinger, dynamics, for two (or more) systems, widely separated from each other and not interacting with each other, evolution of each system is independent, and we can employ separate time variables for each system. The Schr\"odinger dynamics do not require any notion of synchronization of the time variables.  This is a feature that Bell calls \emph{relative time translation invariance}.  The question he addressed is: for systems like that, does the GRW theory retain relative time translation invariance, or do its probabilistic laws require there to be a matter of fact about the temporal order of events at a distance?  Bell's verdict: the probability of any sequence of GRW jumps is relative time translation invariant.
\begin{quote}
I am particularly struck by the fact that the model is as Lorentz invariant as it could be in the
nonrelativistic version.  It takes away the ground of my fear that any exact formulation of quantum mechanics must conflict with fundamental Lorentz invariance (\citealt[p. 14]{BellQJ}, in \citealt[p. 209]{BellBook}).
\end{quote}

In his Trieste lecture, delivered in November 1989,  Bell discussed the prospects for a genuinely relativistic version of  a dynamical collapse theory, and concluded that the difficulties encountered by Ghirardi, Grassi, and Pearle in producing a genuinely relativistic version of CSL  a theory that would be ``Lorentz invariant, not just for all practical purposes but deeply, in the sense of Einstein, eliminating
entirely any privileged reference system from the theory'' \citep[p. 2931]{BellTrieste}, were ``Second-Class Difficulties,'' technical difficulties, and not deep conceptual ones (see also \citealt[p. 25]{BellExactQM}).

A relativistic collapse theory, if it is to yield even an approximation to the usual quantum probabilities, will include correlations between events at spacelike separation that violate Bell inequalities.  The Bell inequalities can be derived from the condition that correlations be \emph{locally explicable}.  This has two parts: that the correlations be explicable in a certain sense, and the explanation be local.  The explicability condition was taken by Bell to be the condition that  correlations  between events that are not in a direct cause-effect relation with each other be attributable to a common cause (see \citealt[p. C2–55]{BellSox}, in \citealt[p. 152]{BellBook}).  This condition is violated in relativistic collapse theories, which involve probabilistic correlations between spacelike separated events that are \emph{not} attributable in the usual way to events in their common past.  Unlike cause-and-effect relations as usually conceived, the relation of probabilistic correlation between these distant events is symmetric, and does not require a temporal order between the events.  For this reason, it is misleading to assimilate this relation to the causal relation and refer to it as nonlocal ``influence'' or ``action at a distance.'' See \citet{BellPaper} for further discussion.

As we shall see, there are difficulties associated with constructing a fully relativistic collapse theory. These difficulties arise, not because a collapse theory cannot be formulated in the absence of a distinguished relation of distant simultaneity, but rather from the problem of constructing a relativistic collapse theory with a stable vacuum state.  In order to see how all this works, we start with a few words about collapse theories in general.

\subsection{Stochastic quantum state evolution}\label{SQSE} It is generally accepted that any  physically realizable change of the state of a quantum system must be a \emph{completely positive} mapping of the system's state space into itself.  Any such mapping can be represented by a set $\{K_i\}$ of operators, such that the density operator $\rho$ representing the state undergoes the change
\begin{equation}
\rho \rightarrow \sum_i K_i \: \rho \:K_i^\dag,
\end{equation}
where
\begin{equation}
\sum_i K_i^\dag K_i \leq \mathds{1}.
\end{equation}
If $\sum_i K_i^\dag K_i = \mathds{1}$, the operation is called a \emph{nonselective operation}; if   $\sum_i K_i^\dag K_i < \mathds{1}$, it is \emph{selective}.  Such a representation of a completely positive mapping of the state space into itself is called a \emph{Kraus representation} of the mapping; and the operators $\{K_i\}$, \emph{Kraus operators}.

Selective operations reduce the trace-norm of the density operator, but this is not an issue, as normalization is only a convention. With an unnormalized density operator, we compute the expectation value of an observable represented by an operator $A$ via
\begin{equation}
\langle A \rangle_\rho = \mbox{Tr}(\rho A)  / \mbox{Tr}(\rho).
\end{equation}

In the simplest case, the set of Kraus operators is a singleton, and the evolution is deterministic; this includes the case of unitary evolution.  However, one can also consider stochastic processes. Suppose that we has a set of operators $\{ K_i \}$, such  that the state vector undergoes a stochastic transition of the following form. For some $i$,
\begin{equation}\label{trans}
\ket{\psi} \Rightarrow \ket{\psi'} = K_i \ket{\psi}/ \nrm{K_i \ket{\psi}} ,
\end{equation}
with the probability for which transition it undergoes given by
\begin{equation}
p_i = \nrm{K_i \ket{\psi}}^2.
\end{equation}
Since these probabilities must sum to unity  for every vector $\ket{\psi}$, we must have $\sum_i K_i^\dag K_i = \mathds{1}$.  For each $i$, we have a selective operation.  One of these yields the actual state.  The transition to the mixture of these candidate states, which corresponds to the proposition that some one of these transitions has occurred, without specification of which, is given by weighted mean of these.
\begin{equation}
\rho \rightarrow \bar{\rho'} = \sum_i p_i \: \frac{K_i \rho K_i^\dag}{\mbox{Tr}(K_i \rho K_i^\dag)}.
\end{equation}

In (\ref{trans}), we indexed the possible state transitions via a discrete index $i$.  We can also consider stochastic processes of a more general sort.  Let $\langle \Gamma, \mc{F}, \mu \rangle$ be a probability space.\footnote{That is, $\Gamma$ is some set, $\mc{F}$ is a $\sigma$-algebra of subsets of $\Gamma$ containing both the empty set $\emptyset$ and $\Gamma$ itself, to be regarded as the measurable subsets of $\Gamma$,  and $\mu$ is a non-negative countably additive set function on $\mc{F}$ with $\mu(\Gamma) = 1$.}  Suppose that  we have a family of operators $\{K_\gamma\}$, for $\gamma \in \Gamma$,  such that
\begin{equation}
\int_\Gamma K^\dag_\gamma K_\gamma \, d\mu(\gamma) = \mathds{1}.
\end{equation}
These can serve as the operators that induce our state transitions.

This gives us a rather general schema for a stochastic process in a Hilbert space.    For the moment, we will presume that we have a unique global time function; extension to a relativistic spacetime will be considered in the next subsection.

Any theory satisfying the following conditions will give us stochastic evolution of the state vector.
\begin{enumerate}
\item    We assume a measure space $\langle \Gamma, \mc{F}, \mu \rangle$.  The elements of $\Gamma$ are to be thought of as possible complete histories, specifying events for all times.  For each time $t$  there is a $\sigma$-algebra $\mc{F}_t$, whose elements are to be thought of as sets of  possible histories up to time $t$.  For $t < s$, we will have $\mc{F}_t \subseteq  \mc{F}_s \subseteq  \mc{F}$.
\item Given a history up until time $t$, and a state vector $\ket{\psi(t)}$, the state at a later time $t + s$ is a random variable, defined as follows.
    \begin{enumerate}
    \item We assume a family of operators $K_\gamma(s; t)$, which depend only on events up until time $s$ (that, is, if two histories $\gamma, \gamma'$ agree up to time $s$, then $K_\gamma(s; t) = K_{\gamma'}(s; t)$), such that
        \[
        \int_\Gamma K^\dag_\gamma(s; t) K_\gamma(s; t) \, d\mu = \mathds{1}.
        \]
    \item For some $\gamma$,
    \[
    \ket{\psi(t + s)} = \frac{K_\gamma(s; t) \ket{\psi(t)}}{\nrm{K_\gamma(s; t) \ket{\psi(t)}}}.
    \]
    \item The probability distribution for the realized outcome is given by
    \[
    \mbox{Prob}(\gamma \in F) = \int_F \nrm{K_\gamma(s; t) \ket(\psi(t)}^2 \, d\mu.
    \]
    for any $F$ in $\mc{F}_s$.
    \end{enumerate}
\end{enumerate}

On a theory of this sort, given a state  $\ket{\psi(t_0)}$, for any later time $t_1 = t_0 +  s$ there will be an actual state vector $\ket{\psi(t_1)}$, and its corresponding pure-state density operator
\begin{equation}
\rho(t_1) = \ket{\psi(t_1)} \bra{\psi(t_1)}.
\end{equation}

There will also be what may be called an \emph{ensemble}, or \emph{mean} density operator, which is a weighted average of the various possibilities for $\rho(t_1)$, weighted by their respective probabilities.
\begin{equation}\label{KrausEnsemble}
\bar{\rho}(t_1; t_0) =  \int_{\Gamma} K_\gamma(t_1; t_0) \, \rho(t_0) \,  K^\dag_\gamma(t_1; t_0) \, d\mu.
\end{equation}

\subsection{Collapse theories in Minkowski spacetime}

\subsubsection{Causality conditions}

It is useful to work within what may be called the \emph{stochastic Tomonaga-Schwinger picture}.  We start with field operators that are solutions of the standard Heisenberg-picture equations (these may be free or interacting fields).  We associate with each spacelike Cauchy surface $\sigma$ a state vector $\ket{\psi(\sigma)}$. Our theory should deliver a stochastic evolution of $\ket{\psi(\sigma)}$ to $\ket{\psi(\sigma')}$ whenever $\sigma'$ is nowhere to the past of $\sigma$.  This will go much as it did in \S \ref{SQSE}.

\begin{enumerate}
\item    We assume a measure space $\langle \Gamma, \mc{F}, \mu \rangle$.  The elements of $\Gamma$ are to be thought of as possible complete histories, specifying events for all times.  For each spacelike Cauchy surface  $\sigma$  there is a $\sigma$-algebra $\mc{F}_\sigma$, whose elements are to be thought of as sets of histories to the  past of $\sigma$.  For two Cauchy surfaces $\sigma, \tau$, with $\tau \preceq \sigma$, we will have $\mc{F}_\tau \subseteq  \mc{F}_\sigma \subseteq  \mc{F}$.
\item Given two Cauchy surfaces $\sigma, \tau$, with $\tau \preceq \sigma$, let $\delta$ be the spacetime region between them.  Given  a state vector $\ket{\psi(\tau)}$, the state vector $\ket{\psi(\sigma)}$ is a random variable, defined as follows.
    \begin{enumerate}
    \item We assume a family of operators $\{K_\gamma(\delta)\}$, which depend only on events in the past of  $\sigma$ (that, is, if two histories $\gamma, \gamma'$ agree on events in the past of $\sigma$  then $K_\gamma(\delta) = K_{\gamma'}(\delta)$), such that
        \[
        \int_\Gamma K^\dag_\gamma(\delta) K_\gamma(\delta) \, d\mu = \mathds{1}.
        \]
        We call these \emph{evolution operators} for the region $\delta$.
    \item For some $\gamma$,
    \[
    \ket{\psi(\sigma)} = \frac{K_\gamma(\delta) \ket{\psi(\tau)}}{\nrm{K_\gamma(\delta) \ket{\psi(\tau)}}}.
    \]
    \item The probability distribution for the realized outcome is given by
    \[
    \mbox{Prob}(\gamma \in F) = \int_F \nrm{K_\gamma(\delta) \ket{\psi(\tau)}}^2 \, d\mu.
    \]
    for any $F$ in $\mc{F}_\sigma$.
    \end{enumerate}
\end{enumerate}
Here, again, we can define a mean density operator for the state on $\sigma$, given a state vector on $\tau$, with $\tau \subseteq \sigma$.

\begin{equation}\label{KrausEnsembleR}
\bar{\rho}(\sigma; \tau) =  \int_{\Gamma} K_\gamma(\delta) \rho(\tau) K^\dag_\gamma(\delta) \, d\mu.
\end{equation}

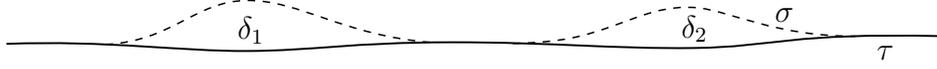
\begin{figure}[h]
\begin{centering}
\ifx\du\undefined
  \newlength{\du}
\fi
\setlength{\du}{8\unitlength}
\begin{tikzpicture}
\pgftransformxscale{1.000000}
\pgftransformyscale{-1.000000}
\definecolor{dialinecolor}{rgb}{0.000000, 0.000000, 0.000000}
\pgfsetstrokecolor{dialinecolor}
\definecolor{dialinecolor}{rgb}{1.000000, 1.000000, 1.000000}
\pgfsetfillcolor{dialinecolor}
\pgfsetlinewidth{0.127000\du}
\pgfsetdash{}{0pt}
\pgfsetdash{}{0pt}
\pgfsetmiterjoin
\pgfsetbuttcap
{
\definecolor{dialinecolor}{rgb}{0.000000, 0.000000, 0.000000}
\pgfsetfillcolor{dialinecolor}
\definecolor{dialinecolor}{rgb}{0.000000, 0.000000, 0.000000}
\pgfsetstrokecolor{dialinecolor}
\pgfpathmoveto{\pgfpoint{0.863496\du}{7.906910\du}}
\pgfpathcurveto{\pgfpoint{7.613500\du}{7.656910\du}}{\pgfpoint{12.324900\du}{8.406610\du}}{\pgfpoint{16.858800\du}{8.382320\du}}
\pgfpathcurveto{\pgfpoint{21.392600\du}{8.358030\du}}{\pgfpoint{23.691800\du}{7.808850\du}}{\pgfpoint{30.408500\du}{7.800520\du}}
\pgfpathcurveto{\pgfpoint{37.125100\du}{7.792190\du}}{\pgfpoint{41.101000\du}{8.249700\du}}{\pgfpoint{46.558800\du}{8.182320\du}}
\pgfpathcurveto{\pgfpoint{52.016600\du}{8.114940\du}}{\pgfpoint{52.655400\du}{7.146250\du}}{\pgfpoint{63.155400\du}{7.396250\du}}
\pgfusepath{stroke}
}
\definecolor{dialinecolor}{rgb}{0.000000, 0.000000, 0.000000}
\pgfsetstrokecolor{dialinecolor}
\node[anchor=west] at (51.199700\du,6\du){$\sigma$};
\pgfsetlinewidth{0.100000\du}
\pgfsetdash{{1.000000\du}{1.000000\du}}{0\du}
\pgfsetdash{{0.500000\du}{0.500000\du}}{0\du}
\pgfsetmiterjoin
\pgfsetbuttcap
{
\definecolor{dialinecolor}{rgb}{0.000000, 0.000000, 0.000000}
\pgfsetfillcolor{dialinecolor}
\definecolor{dialinecolor}{rgb}{0.000000, 0.000000, 0.000000}
\pgfsetstrokecolor{dialinecolor}
\pgfpathmoveto{\pgfpoint{7.441250\du}{7.942050\du}}
\pgfpathcurveto{\pgfpoint{11.949100\du}{7.729920\du}}{\pgfpoint{13.358000\du}{5.059870\du}}{\pgfpoint{16.849700\du}{5.018200\du}}
\pgfpathcurveto{\pgfpoint{20.341300\du}{4.976530\du}}{\pgfpoint{23.249700\du}{7.368200\du}}{\pgfpoint{29.199700\du}{7.818200\du}}
\pgfusepath{stroke}
}
\pgfsetlinewidth{0.100000\du}
\pgfsetdash{{0.500000\du}{0.500000\du}}{0\du}
\pgfsetdash{{0.500000\du}{0.500000\du}}{0\du}
\pgfsetmiterjoin
\pgfsetbuttcap
{
\definecolor{dialinecolor}{rgb}{0.000000, 0.000000, 0.000000}
\pgfsetfillcolor{dialinecolor}
\definecolor{dialinecolor}{rgb}{0.000000, 0.000000, 0.000000}
\pgfsetstrokecolor{dialinecolor}
\pgfpathmoveto{\pgfpoint{34.514600\du}{7.880180\du}}
\pgfpathcurveto{\pgfpoint{37.929000\du}{7.806350\du}}{\pgfpoint{39.304700\du}{7.374590\du}}{\pgfpoint{41.154700\du}{6.724590\du}}
\pgfpathcurveto{\pgfpoint{43.004700\du}{6.074590\du}}{\pgfpoint{44.699700\du}{5.268200\du}}{\pgfpoint{46.899700\du}{5.518200\du}}
\pgfpathcurveto{\pgfpoint{49.099700\du}{5.768200\du}}{\pgfpoint{51.971400\du}{7.400000\du}}{\pgfpoint{57.300000\du}{7.400000\du}}
\pgfusepath{stroke}
}
\definecolor{dialinecolor}{rgb}{0.000000, 0.000000, 0.000000}
\pgfsetstrokecolor{dialinecolor}
\node[anchor=west] at (15.5\du,7\du){$\delta_1$};
\definecolor{dialinecolor}{rgb}{0.000000, 0.000000, 0.000000}
\pgfsetstrokecolor{dialinecolor}
\node[anchor=west] at (45\du,6.9\du){$\delta_2$};
\definecolor{dialinecolor}{rgb}{0.000000, 0.000000, 0.000000}
\pgfsetstrokecolor{dialinecolor}
\node[anchor=west] at (58.000000\du,8.5\du){$\tau$};
\end{tikzpicture}\caption{Spacelike separated regions between two spacelike hypersurfaces.}\label{commute}
\end{centering}
\end{figure}

Consider two Cauchy surfaces, $\tau$, $\sigma$, which coincide everywhere except on the boundaries of two bounded regions $\delta_1$, $\delta_2$, where $\sigma$ is to the future of $\tau$ (see Figure \ref{commute}).  We must have a unique law of stochastic evolution from $\tau$ to $\sigma$, through $\delta_1 \cup \delta_2$, and this should coincide with the composition of the evolution through $\delta_1$ and the evolution through $\delta_2$, in either order.  The necessary and sufficient condition for this is,
\begin{enumerate}[resume]
\item \label{LocalEv} Evolution operators  corresponding to spacelike separated regions commute.
\end{enumerate}

One difference between a theories of this sort and a theory involving  deterministic, unitary evolution, is that, for calculating probabilities of outcomes of experiments performed in some spacelike slice $\alpha$ that is common to two Cauchy surfaces  $\tau$, $\sigma$, it can make a difference whether the calculation is made using $\ket{\psi(\tau)}$ or $\ket{\psi(\sigma)}$. It is this feature that has given rise to concerns that  a relativistic collapse theory must either be incoherent or exhibit  radical  contextualism about possessed properties. These concerns are misplaced; more on this, below, in subsection \ref{ontology}. However, it \emph{would} be problematic if probabilities for events in $\alpha$ differed depending on whether one used $\ket{\psi(\tau)}$ or $\bar{\rho}(\sigma)$, which is the probabilistically weighted mixture of various possibilities for $\ket{\psi(\sigma)}$, given the state on $\tau$.  Moreover, if the evolution from $\tau$ to $\sigma$, through a region $\delta$, spacelike separated from $\alpha$, depended on a choice of experiments to be made, then, if different choices yielded different mixtures $\bar{\rho}(\sigma)$, the choice of experiment could be used for superluminal signalling.  We, therefore, require $\bar{\rho}(\sigma)$ to yield the same probabilities for outcomes of experiments as $\ket{\psi(\tau)}$, for any  allowable evolution through $\delta$.   The necessary and sufficient condition for this is \citep{FixedPoints},
\begin{enumerate}[resume]
\item \label{NoSig} Evolution operators pertaining to a region $\delta$ commute with all operators representing observables at spacelike separation from $\delta$.
\end{enumerate}
The conditions \ref{LocalEv}, \ref{NoSig} ensure compatibility with relativistic causal structure.

\subsubsection{Collapse theory ontology}\label{ontology}  If the region between two Cauchy surfaces  $\tau$, $\sigma$ is a bounded spacetime region $\delta$, then the transition from the state on $\tau$ to the state on $\sigma$ can be attributed to events within $\delta$.  Quantum states, on the other hand, are in some sense ``nonlocal''---a better term is \emph{nonseparable}.  They cannot be regarded as supervening on local matters of fact.  A quantum state encodes, not merely local matters of fact, but also correlations between distant regions.

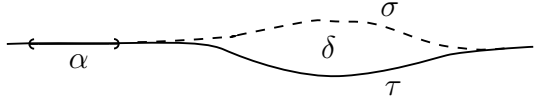
\begin{figure}[h]
\begin{centering}
\ifx\du\undefined
  \newlength{\du}
\fi
\setlength{\du}{10\unitlength}
\begin{tikzpicture}
\pgftransformxscale{1.000000}
\pgftransformyscale{-1.000000}
\definecolor{dialinecolor}{rgb}{0.000000, 0.000000, 0.000000}
\pgfsetstrokecolor{dialinecolor}
\definecolor{dialinecolor}{rgb}{1.000000, 1.000000, 1.000000}
\pgfsetfillcolor{dialinecolor}
\pgfsetlinewidth{0.100000\du}
\pgfsetdash{}{0pt}
\pgfsetdash{}{0pt}
\pgfsetmiterjoin
\pgfsetbuttcap
{
\definecolor{dialinecolor}{rgb}{0.000000, 0.000000, 0.000000}
\pgfsetfillcolor{dialinecolor}
\definecolor{dialinecolor}{rgb}{0.000000, 0.000000, 0.000000}
\pgfsetstrokecolor{dialinecolor}
\pgfpathmoveto{\pgfpoint{1.500000\du}{12.850000\du}}
\pgfpathcurveto{\pgfpoint{1.798800\du}{12.850000\du}}{\pgfpoint{1.875000\du}{12.787500\du}}{\pgfpoint{5.525000\du}{12.812500\du}}
\pgfpathcurveto{\pgfpoint{9.175000\du}{12.837500\du}}{\pgfpoint{11.800000\du}{12.600000\du}}{\pgfpoint{13.050000\du}{13.150000\du}}
\pgfpathcurveto{\pgfpoint{14.300000\du}{13.700000\du}}{\pgfpoint{17.050000\du}{14.700000\du}}{\pgfpoint{19.400000\du}{14.550000\du}}
\pgfpathcurveto{\pgfpoint{21.750000\du}{14.400000\du}}{\pgfpoint{24.016700\du}{13.675000\du}}{\pgfpoint{24.900000\du}{13.450000\du}}
\pgfpathcurveto{\pgfpoint{25.783300\du}{13.225000\du}}{\pgfpoint{29.201200\du}{13.000000\du}}{\pgfpoint{29.500000\du}{13.000000\du}}
\pgfusepath{stroke}
}
\pgfsetlinewidth{0.100000\du}
\pgfsetdash{{1.000000\du}{1.000000\du}}{0\du}
\pgfsetdash{{0.500000\du}{0.500000\du}}{0\du}
\pgfsetmiterjoin
\pgfsetbuttcap
{
\definecolor{dialinecolor}{rgb}{0.000000, 0.000000, 0.000000}
\pgfsetfillcolor{dialinecolor}
\definecolor{dialinecolor}{rgb}{0.000000, 0.000000, 0.000000}
\pgfsetstrokecolor{dialinecolor}
\pgfpathmoveto{\pgfpoint{8.450000\du}{12.750000\du}}
\pgfpathcurveto{\pgfpoint{9.495800\du}{12.750000\du}}{\pgfpoint{12.375000\du}{12.575000\du}}{\pgfpoint{13.400000\du}{12.450000\du}}
\pgfpathcurveto{\pgfpoint{14.425000\du}{12.325000\du}}{\pgfpoint{11.650000\du}{12.800000\du}}{\pgfpoint{15.150000\du}{12.100000\du}}
\pgfpathcurveto{\pgfpoint{18.650000\du}{11.400000\du}}{\pgfpoint{17.700000\du}{11.600000\du}}{\pgfpoint{19.000000\du}{11.650000\du}}
\pgfpathcurveto{\pgfpoint{20.300000\du}{11.700000\du}}{\pgfpoint{20.550000\du}{11.500000\du}}{\pgfpoint{22.650000\du}{12.300000\du}}
\pgfpathcurveto{\pgfpoint{24.750000\du}{13.100000\du}}{\pgfpoint{25.200000\du}{13.000000\du}}{\pgfpoint{25.950000\du}{13.100000\du}}
\pgfpathcurveto{\pgfpoint{26.700000\du}{13.200000\du}}{\pgfpoint{27.150000\du}{13.150000\du}}{\pgfpoint{28.350000\du}{13.050000\du}}
\pgfusepath{stroke}
}
\pgfsetlinewidth{0.100000\du}
\pgfsetdash{}{0pt}
\pgfsetdash{}{0pt}
\pgfsetbuttcap
{
\definecolor{dialinecolor}{rgb}{0.000000, 0.000000, 0.000000}
\pgfsetfillcolor{dialinecolor}
}
\definecolor{dialinecolor}{rgb}{0.000000, 0.000000, 0.000000}
\pgfsetstrokecolor{dialinecolor}
\draw (3.199998\du,12.848582\du)--(6.950002\du,12.838918\du);
\pgfsetlinewidth{0.100000\du}
\pgfsetdash{}{0pt}
\pgfsetmiterjoin
\pgfsetbuttcap
\definecolor{dialinecolor}{rgb}{0.000000, 0.000000, 0.000000}
\pgfsetstrokecolor{dialinecolor}
\pgfpathmoveto{\pgfpoint{2.949361\du}{12.599228\du}}
\pgfpatharc{270}{89}{0.250000\du and 0.250000\du}
\pgfusepath{stroke}
\definecolor{dialinecolor}{rgb}{0.000000, 0.000000, 0.000000}
\pgfsetstrokecolor{dialinecolor}
\draw (3.199998\du,12.848582\du)--(2.700000\du,12.849871\du);
\pgfsetlinewidth{0.100000\du}
\pgfsetdash{}{0pt}
\pgfsetmiterjoin
\pgfsetbuttcap
\definecolor{dialinecolor}{rgb}{0.000000, 0.000000, 0.000000}
\pgfsetstrokecolor{dialinecolor}
\pgfpathmoveto{\pgfpoint{7.200626\du}{13.088272\du}}
\pgfpatharc{90}{-90}{0.250000\du and 0.250000\du}
\pgfusepath{stroke}
\definecolor{dialinecolor}{rgb}{0.000000, 0.000000, 0.000000}
\pgfsetstrokecolor{dialinecolor}
\draw (6.950002\du,12.838918\du)--(7.450000\du,12.837629\du);
\definecolor{dialinecolor}{rgb}{0.000000, 0.000000, 0.000000}
\pgfsetstrokecolor{dialinecolor}
\node[anchor=west] at (4.2\du,13.787500\du){$\alpha$};
\definecolor{dialinecolor}{rgb}{0.000000, 0.000000, 0.000000}
\pgfsetstrokecolor{dialinecolor}
\node[anchor=west] at (21\du,15.2\du){$\tau$};

\definecolor{dialinecolor}{rgb}{0.000000, 0.000000, 0.000000}
\pgfsetstrokecolor{dialinecolor}
\node[anchor=west] at (17.625000\du,13\du){$\delta$};
\definecolor{dialinecolor}{rgb}{0.000000, 0.000000, 0.000000}
\pgfsetstrokecolor{dialinecolor}
\node[anchor=west] at (20.775000\du,11\du){$\sigma$};
\end{tikzpicture}\caption{A region common to two spacelike hypersurfaces.}\label{alphastate}
\end{centering}
\end{figure}

This is reflected in the way that a collapse affects the global state. Let $\tau$ and $\sigma$ be two spacelike hypersurfaces, that coincide everywhere except on the boundaries of a bounded region $\delta$, where $\sigma$ lies to the future of $\tau$.  Let $\alpha$ be a bounded region that lies in both $\tau$ and $\sigma$, and hence is at spacelike separation from $\delta$ (see Figure \ref{alphastate}).  The two states can differ on probabilities assigned to observables pertaining to $\alpha$.  For example, consider the familiar EPR-Bohm scenario.  Suppose the state on $\tau$ involves an entangled  pair of spin-$1/2$ particles, one of which is located (to the extent that it can be located)  within $\alpha$, and suppose that in $\delta$ an experiment is performed on the other one, yielding, on $\sigma$, a near eigenstate of spin in some direction ${a}$ that is correlated by the state on $\tau$ with spin of the other particle in direction ${b}$.  Then the state on $\tau$ is a near-eigenstate of  ${b}$-spin for the second particle.  Thus, the two states $\ket{\psi(\tau)}$ and $\ket{\psi(\sigma)}$ differ on probabilities assigned to experiments performed in the forward domain of dependence of $\alpha$.
Relativistic causality precludes events occurring in a region $\delta$ from affecting  \emph{local} beables pertaining to regions at spacelike separation from $\delta$.   If one thought that the differing  probabilities assigned to observables pertaining to $\alpha$ by the states $\ket{\psi(\tau)}$ and $\ket{\psi(\sigma)}$  were {local beables}, this would give rise to a serious worry about the conceptual coherence of a relativistic collapse theory (see, \emph{e.g.} \citealt[p. 209]{Maudlin1} for worries along these lines).

But the reduced states of $\alpha$ obtained from the global states $\ket{\psi(\tau)}$ and  $\ket{\psi(\sigma)}$ are not obviously local matters of fact about $\alpha$, and, indeed, what these reduced states are depends on events at spacelike separation from $\alpha$. What \emph{can}, and \emph{should} be regarded as the intrinsic state of the spacetime region $\alpha$, that is, its state stripped of implicit references to goings-on at a distance from $\alpha$, is the past light cone state.  This is the limiting state obtained from a sequence of spacelike hypersurfaces that converge on the past light cone of $\alpha$. This state will reflect collapse events to the past of $\alpha$, but not collapse events  at spacelike separation.

This has implications for what the local beables of the theory could be.  Ghirardi and Grassi (\citeyear[p. 419]{GG94}) proposed a criterion for property attribution in the context of relativistic collapse theories.  If $A$ is a local observable with support in $\alpha$, a system possesses a property $A = a$ when the expectation value of $P_a$, the projection onto the subspace $A = a$, is extremely close to 1, evaluated with respect to the past light cone state of $\alpha$.

Ghirardi, Grassi, and Benatti (\citeyear{GGB95}) proposed, as an appropriate local beable for collapse theories, a smeared mass density, with the smearing taken over lengths scales large compared to the atomic scale but small on a macroscopic scale.  For a relativistic theory, the corresponding quantity would be a smeared  energy-momentum density, whose $00$ component is mass-energy density, which, in a nonrelativistic context, is dominated by the rest mass.   Combining the choice of mass density as local beable with the past light cone criterion for property attribution, the mass density that in a relativistic context is taken to be a local beable in a region $\alpha$ is  the mass density defined by the state on the past light cone of $\alpha$ \citep{Tumulka2007,MatterDensity}. See \citet{BellPaper,CollapseOntology,RelCollapseOntology} for further discussion of ontological issues in collapse theories.

\subsubsection{A problem: stability of the vacuum} There is a difficulty with constructing a physically sensible collapse theory that respects the full symmetries of Minkowski spacetime.  This has to do with the stability of the vacuum. A relativistic collapse theory must have a stable vacuum, which is not achievable via a straightforward extension of CSL to a relativistic quantum field theory.

If the theory were to produce excitations of the vacuum, it would have to furnish a probability distribution over possible excitations. The difficulty is the nonexistence of a bounded Lorentz-invariant measure over possible excitations.

This is readily seen in  the simplest case, a free theory.\footnote{This paragraph is based on the discussion in \citet[p. 2]{Pearle2015}.} Let $\Delta$ be some subset of momentum space, and suppose that our collapse theory yields a certain probability $p(\Delta)$ per unit time of producing from the vacuum a particle with momentum  in $\Delta$.  Lorentz invariance requires that there be the same probability for producing a particle with its momentum in any Lorentz boost of this volume.

This requirement cannot be satisfied.  The problem is that there is no bounded invariant measure on the mass-shell consisting of all four-momenta of a given magnitude $m$. Any probability measure on the mass-shell, therefore, must break Lorentz symmetry.  In the vacuum state, which is a Lorentz invariant state, there is nothing that could be used to define such a measure.  To generalize beyond free theories, the issue is the absence of a bounded Lorentz-invariant measure over the space of possible excitations of the vacuum.

The requirement of a strictly stable vacuum applies only to theories that respect the full set of spacetime symmetries of Minkowski spacetime.  One could, for example, have a theory on a lattice, with relativistic causal structure.  The rate of energy increase could, for such a theory, be kept finite.  Whether it could be kept below the threshold of current observations would depend on the lattice spacing and other parameters of the theory; see Ghirardi, Grassi and Pearle (\citeyear[p. 1297]{GGP1990}) for discussion.

\subsubsection{Impossibility of a relativistic collapse theory using only standard degrees of freedom} Though relativistic causality does not preclude the formulation of a relativistic collapse theory, there is a tension between the requirements of  relativistic causality on a collapse theory and the need for a stable vacuum. It is possible to avoid an outright contradiction, but only at the cost of introducing nonstandard degrees of freedom, that do not obey anything like the usual laws of quantum evolution.

Here's the problem. Suppose we have a collapse theory with a stable vacuum state.  Consider evolution from one surface $\tau$ to another, $\sigma$, through a bounded spacetime region $\delta$.  Stability of the vacuum means  if the state on $\tau$ is the vacuum state,   subsequent evolution is deterministic, in that there is only a single outcome that has nonzero probability, namely, maintenance of the vacuum.  Deterministic evolution starting from vacuum means that, except perhaps for a set with zero probability of being realized,  all collapse operators $\{K_\gamma(\delta) \}$ map the vacuum into multiples of the same vector. It can be shown (see \citealt{CollapseNoGo} for details) that this will also hold for all standard states, where standard states are those that can be obtained by operating on the vacuum with operators of the sort that appear in standard quantum field theories.  That is, for standard quantum field theories, the constraints imposed by relativistic causal structure,  plus the constraint, imposed by Lorentz invariance, of a stable vacuum, are satisfied at the expense of rendering the theory deterministic, and thus incapable of suppressing the sorts of superpositions of macroscopically distinct states that collapse theories are meant to suppress.

This is not a no-go result for relativistic collapse theories \emph{tout court}.  \citet{BedinghamRSRM,BedinghamRSRD} and \citet{Pearle2015} utilize a nonstandard field first introduced by \citet{PearleWays}.  This ``index'' or ``pointer'' field  does not obey anything like an ordinary law of evolution; instead, its value at any point is independent of its value at any other point.    The vacuum state $\ket{\Omega}$  of this theory is the ground state of the standard fields \emph{and} this pointer field. Standard operators operating on the vacuum state leave the pointer field in its ground state.  On Bedingham's theory the pointer field couples to the standard fields, so that, if there are excitations of the standard fields, these produce excitations of the pointer fields.  The theory tends to collapse towards eigenstates of smeared pointer-field operators,  with accompanying collapse towards eigenstates of smeared matter fields.

One might also reject the requirement of having a theory that produces collapses in Minkowski spacetime. \citet{OkonSudarsky2014,OkonSudarsky2016,WeightofCollapse} have proposed, in the context of semi-classical gravity, a modification of the CSL theory on which the collapse rate depends on local curvature. One might consider a modification of this proposal, according to which collapse rate goes to zero as curvature goes to zero \citep{InflationLandscape}.  If there is a viable theory of this sort, it might  satisfy the constraint of having a stable vacuum in Minkowski space, while producing collapse in the presence of matter.

\subsection{Causal Quantum Theory}\label{CQT}  Standard quantum theory predicts correlations between outcomes of spacelike separated experiments that violate Bell inequalities.  The crucial assumption of derivation of  Bell inequalities is that the correlation be \emph{locally explicable}, that is, that they can be explained by reference to conditions in the past lightcones of the experiments, such that a full specification of these conditions screens off the correlations. Experimental demonstrations of violations of Bell inequalities are usually taken to indicate correlations that are not locally explicable are a genuine physical phenomenon (see Myrvold, Genovese, and Shimony \citeyear{sep-bell-theorem} for an overview, and a discussion of the conditions required to reach this conclusion).

There is a loophole, however, identified by \citet{KentCQT}, and called by him the \emph{collapse locality loophole}.  Tests of Bell inequalities that aim to establish nonlocality take care to ensure that the experiments involved take place at spacelike separation from each other, from choice of experimental setting to registration of result. A judgment of whether this condition is satisfied requires, therefore, a judgment of when an experiment is over. Typically it is assumed that this occurs once the result has been permanently recorded in some macroscopically readable device.  However, as Kent observes, on some proposals, the quantum state will remain in a superposition corresponding to distinct outcomes beyond this point.  One such proposal is the suggestion that state reduction takes place only when the uncollapsed state involves a superposition of sufficiently distinct gravitational fields \citep{Diosi87,PenroseNewMind,Penrose96}.  Another is Wigner’s suggestion that conscious awareness of the result is required to induce collapse \citep{WignerMindBody}.

Kent introduces a class of theories, which he calls \emph{causal quantum theories}, on which collapses are localized events, collapse probabilities are conditioned only on events in the collapse's past light cone, and, as a consequence, collapses at spacelike separation from each other are probabilistically independent, conditional on events in their past lightcones.  A theory of that sort differs in its predictions from standard quantum mechanics, but a decisive test requires experiments that are completed at spacelike separation from each other.

On a theory of the sort proposed by Di\'osi and Penrose, on which collapse occurs only when differing results correspond to sufficiently distinct gravitational fields, the condition that the experiments be concluded at spacelike separation had not been enforced in any of the experimental tests of Bell inequalities as of the time of writing (2005). The experiment of \citet{Salart2008} closed the loophole for the particular proposals of Penrose and Di\'osi, though, as \citet{KentTCQT}  pointed out, altering the Penrose-Di\'osi threshold by a few orders of magnitude would render these proposals compatible with the results of this experiment.

No experiment to date has addressed the collapse locality loophole if the collapse condition is taken to be awareness of the result by a conscious observer. See \citet{KentTCQT} for proposals of ways in which causal quantum theory could be subjected to more stringent tests.

\section{Everettian Interpretations}  Everettian interpretations make use of  minimal  machinery: they leave  the usual, unitary dynamics untouched, and add nothing to the quantum state.  The complications arise in connection with  painting a sensible picture of the world with this minimal palette, and with making sense of probabilities or a working substitute for them.

Because of this minimality, worries about compatibility with relativity are also minimal.  If the theory to be interpreted is a relativistic quantum field theory, then there is no question about the dynamics being compatible with relativistic causal structure; it is.  As there are no additional beables, there is no question about whether their dynamics is compatible with relativity. Some proponents of Everettian interpretations have taken this to be one of the chief attractions of interpretations of this sort (see, \emph{e.g.} \citealt{VaidmanPSA94,VaidmanBellGao,sep-qm-manyworlds}).

Given a bounded subset $\alpha$ that is part of many Cauchy surfaces $\sigma,$ $\sigma'$, \emph{etc.}, we can consider the reduced state of $\alpha$,\footnote{Those who want to insist that states are to be assigned only to regions with nonempty interior may take this to mean: the state of the causal domain of dependence of $\alpha$.} evaluated with respect to various Tomonaga-Schwinger state vectors $\ket{\psi(\sigma)}$,  $\ket{\psi(\sigma')}$.  If the evolution from one Cauchy surface to another is unitary, and if the generators of evolutions through regions spacelike separated from $\alpha$ commute with operators representing observables pertaining to $\alpha$, then the reduced state of $\alpha$ will be the same whether calculated from $\ket{\psi(\sigma)}$ or $\ket{\psi(\sigma')}$ or any Cauchy surface containing $\alpha$.  So, in that sense, on any quantum-state monist  theory with deterministic, unitary dynamics, operations performed at a distance have no effect on states of affairs in $\alpha$.

Is there nonlocality, in any sense, in an Everettian theory?  It is not immediately clear how to pose the question.  Unlike, say, the case of additional beables theories, we cannot ask (at least in any straightforward way) whether a choice of a parameter setting at one location affects the outcome of an experiment performed at another, because, on an Everettian theory, talk of \emph{the outcome} of an experiment is unwarranted---experiments do not have unique outcomes. If the usual notions of causal dependence, such as act-outcome correlations, presuppose unique outcomes, then they cannot even be formulated in an Everettian context---and \emph{ipso facto} they cannot be violated. Instead of saying that Everettian theories are causally local, perhaps it would be more apt to say: it is not the case that they are causally nonlocal.

\citet{BrownTimpsonBellGao} have made a stronger claim about Everettian interpretations. Not only is there no dynamical nonlocality, they claim,  there are no nonlocal correlations.  The claim is that, in the Everettian context, talk of correlated outcomes of experiments makes sense only when the outcomes of the distant experiments are compared, at which time both experiments are in the past.  A similar argument has been made of which quantum theories are but a special case, by Brassard and Raymond-Robichaud (\citeyear{BRR2013},\citeyear{BRR2019}).  Analysis of this claim is beyond the scope of this chapter; we note only that compatibility of Everettian interpretations with relativity does not ride on whether a claim of this sort is accepted or rejected.

\section{Non-Realist and Pragmatist Interpretations} There is a long tradition of denying that the ontology of the physical world contains anything corresponding to a quantum state.   In recent years, views of this sort have been championed by advocates of a position called, by its proponents, \emph{Quantum Bayesianism}, or \emph{QBism}  (Caves, Fuchs and Schack \citeyear{CFSQB}; Fuchs, Mermin, and Schack \citeyear{FMSQB}; Fuchs and Schack \citeyear{FSQBGreeks}).  This is one variant of a class of views commonly referred to as $\psi$\emph{-epistemic} views.\footnote{See \citet{HarriganSpekkens} for a discussion.}   A somewhat related, but different, view is the pragmatist view of quantum mechanics defended by Richard \citet{HealeyPragQM,HealeyBridges,HealeyPQR}.  See, also, \citet{sep-quantum-bayesian} for an overview of related views.

For Healey, quantum chances  for events are assigned much as they have been in the discussion, above, of collapse theories.  The chance of an event, say, an outcome of an experiment, is indexed by the spacetime event from which the chance is to be ascribed.  The chance $Ch_x(e_A)$ of event $e_A$ is, in effect, calculated using what in a dynamical collapse theory would be the past light cone state, which incorporates experimental outcomes that have occurred in the past light cone of $x$.  On Healey's view, ``Any agent who accepts quantum theory and is (momentarily) located at space-time point $x$ should match credence in $e_A$ to $Ch_x(e_A)$,'' where $Ch_x(e_A)$ is the chance evaluated conditional on events in the past light cone of the point $x$ (\citealt[p. 179]{HealeyBellGao}).
Healey regards these chances as objective, in the sense that it is an objective fact that these chances are optimal credences for an agent that has epistemic access to all and only the facts in the past light cone of $x$.  But they are not to be thought of as localized physical magnitudes.
\begin{quote}
As we saw, the chance of outcome $e_A$ does not attach to it in virtue solely of its physical description: the \emph{chances} of $e_A$ attach also in virtue of its space-time relations to different space-time locations. Each such location offers the epistemic perspective of a situated agent, even in a world with no such agents. The existence of these chances is independent of the existence of cognizers. But it is only because we are not merely cognizers but physically situated agents that  we have needed to develop a concept of chance tailored to our needs as informationally deprived agents. Quantum chance admirably meets those needs: an omniscient God could describe and understand the physical world without it (\citealt[p. 183]{HealeyBellGao}; see also \citealt[pp. 175--176]{HealeyRevolution}).
\end{quote}
In \citet{HealeyPQR}, Healey argues that this view should be regarded as a form of scientific realism.

In their statement of QBism, Fuchs, Mermin, and Schack (\citeyear{FMSQB}) take a considerably more radical stance.  First, they not only deny that quantum probabilities represent features of physical reality; for them, these probabilities are not objective in any sense---they are ``personal judgments,'' not backed up by objective facts.  Second, the scope of application of quantum mechanics is severely limited.  In ordinary practice, physicists routinely use quantum mechanics to calculate probabilities about what will be observed by other people if they perform certain experiments, whether or not the agent carrying out the calculation will ever be aware of the results (which might, for example, not be realized within the agent's lifetime).  In contrast, Fuchs \emph{et al.}  restrict an agent's assignment of probabilities to propositions concerning the agent's future experience.

The class of propositions to which the agent assigns probabilities all concern events along her own worldline, and therefore, contains no spacelike separated events.
\begin{quote}
when any agent uses quantum mechanics to
calculate``[cor]relations between the manifold aspects of
[her] experience,'' those experiences cannot be space-like
separated. Quantum correlations, by their very nature, refer
only to time-like separated events: the acquisition of experiences
by any single agent. Quantum mechanics, in the QBist
interpretation, cannot assign correlations, spooky or otherwise,
to space-like separated events, since they cannot be
experienced by any single agent. Quantum mechanics is thus
explicitly local in the QBist interpretation.

And that's all there is to it (Fuchs, Mermin, and Schack, \citeyear{FMSQB}, pp. 750--751).
\end{quote}
There is no action at a distance, on this view, because, there are no distances between events that the agent assigns probabilities to.  The spacetime of events to which the theory is applied is, in effect, a one-dimensional timelike worldline (or else a narrow world-tube). There is no superluminal signalling, because there is no signalling, of any kind (unless writing a note to your future self is said to be sending a signal to your future self).

Approaches that deny the reality of quantum states bypass some of the problems faced by realist approaches by limiting the scope of what is taken to be physical.  In my opinion, however, such approaches must  be regarded as promissory notes.  If one denies that quantum states represent anything in physical reality, this raises the question of what \emph{does} constitute physical reality.  It does not seem, for example, that QBists wish to deny that distant places and other agents exists; they merely claim that such things are outside the scope of legitimate application of quantum mechanics.  This raises the question of what physics \emph{does} apply to them.  If one says nothing at all, then ipso facto one says nothing that conflicts with relativity.  The question still remains whether whatever it is that \emph{does} exist can coexist  peacefully with relativity.

\section{Conclusion} There is a temptation to say that, because of quantum nonlocality, as revealed by Bell's theorem, that quantum theory is flatly incompatible with relativity.  As we have seen, things are not so simple.   It is only additional-beables theories, such as the de Broglie-Bohm theory and related theories, that must violate relativistic causal structure (at least, as long as the additional beables are local beables).  Collapse theories can respect relativistic causal structure; unlike the additional-beables theories, they need no distinguished relation of distant simultaneity.  Full relativistic invariance, however, is possible only at the cost of introducing nonstandard degrees of freedom.  For Everettian theories, relativity poses no special problem. For non-realist interpretations, it is difficult to render a verdict, as proponents of such views are usually less clear about what they {do} think \emph{is} in the world than about what they think is \emph{not}.

\section{Acknowledgment}
Support for this research was provided by Graham and Gale Wright, who sponsor the Graham and Gale Wright Distinguished Scholar Award at the University of Western Ontario.

\bibliographystyle{chicago}

\end{document}